\documentclass[twocolumn,aps,prb,10pt,superscriptaddress,longbibliography]{revtex4-2}

\usepackage{amssymb}
\usepackage{amsmath}
\usepackage{multirow}
\usepackage{glossaries}
\usepackage{graphicx}
\usepackage{dcolumn}
\usepackage{bm}
\usepackage{subfigure}
\usepackage[usenames,dvipsnames,svgnames]{xcolor}
\usepackage{hyperref}
\hypersetup{
pdfnewwindow=true,      
colorlinks=true,        
linkcolor=Blue,         
citecolor=Blue,         
filecolor=Blue,         
urlcolor=Blue           
}
\usepackage{listings} 

\newacronym{adf}{ADF}{angular distribution function}
\newacronym{ann}{ANN}{artificial neural network}
\newacronym{emd}{EMD}{equilibrium molecular dynamics}
\newacronym{dft}{DFT}{density functional theory}
\newacronym{dp}{DP}{deep potential}
\newacronym{gap}{GAP}{Gaussian approximation potential}
\newacronym{hcacf}{HCACF}{heat current autocorrelation function}
\newacronym{hnemd}{HNEMD}{homogeneous nonequilibrium molecular dynamics}
\newacronym{md}{MD}{molecular dynamics}
\newacronym{mlp}{MLP}{machine-learned potential}
\newacronym{nep}{NEP}{neuroevolution potential}
\newacronym{nist}{NIST}{national institute of standards and technology}
\newacronym{nqe}{NQE}{nuclear quantum effect}
\newacronym{rdf}{RDF}{radial distribution function}
\newacronym{rmse}{RMSE}{root mean square error}
\newacronym{scan}{SCAN}{strongly constrained and appropriately normed}
\newacronym{snes}{SNES}{separable natural evolution strategy}
\newacronym{vdos}{VDOS}{vibrational density of states}

\begin{document}

\title{Accurate prediction of heat conductivity of water by a neuroevolution potential}

\author{Ke Xu}
 \affiliation{Department of Physics, Research Institute for Biomimetics and Soft Matter, Jiujiang Research Institute and Fujian Provincial Key Laboratory for Soft Functional Materials Research, Xiamen University, Xiamen 361005, P. R. China.}
\author{Yongchao Hao}
 \affiliation{Department of Physics, Research Institute for Biomimetics and Soft Matter, Jiujiang Research Institute and Fujian Provincial Key Laboratory for Soft Functional Materials Research, Xiamen University, Xiamen 361005, P. R. China.}
\author{Ting Liang}
 \affiliation{Department of Electronic Engineering and Materials Science and Technology Research Center, The Chinese University of Hong Kong, Shatin, N.T., Hong Kong SAR, 999077, P. R. China}
\author{Penghua Ying}
\email{hityingph@163.com}
\affiliation{Department of Physical Chemistry, School of Chemistry, Tel Aviv University, Tel Aviv, 6997801, Israel}
\author{Jianbin Xu}
 \affiliation{Department of Electronic Engineering and Materials Science and Technology Research Center, The Chinese University of Hong Kong, Shatin, N.T., Hong Kong SAR, 999077, P. R. China}
\author{Jianyang Wu}
 \email{jianyang@xmu.edu.cn}
 \affiliation{Department of Physics, Research Institute for Biomimetics and Soft Matter, Jiujiang Research Institute and Fujian Provincial Key Laboratory for Soft Functional Materials Research, Xiamen University, Xiamen 361005, P. R. China.}
 \affiliation{NTNU Nanomechanical Lab, Norwegian University of Science and Technology (NTNU), Trondheim 7491, Norway}
\author{Zheyong Fan}
 \email{brucenju@gmail.com}
 \affiliation{College of Physical Science and Technology, Bohai University, Jinzhou 121013, P. R. China}

\date{\today}

\begin{abstract}
We propose an approach that can accurately predict the heat conductivity of liquid water. On the one hand, we develop an accurate machine-learned potential based on the neuroevolution-potential approach that can achieve quantum-mechanical accuracy at the cost of empirical force fields. On the other hand, we combine the Green-Kubo method and the spectral decomposition method within the homogeneous nonequilibrium molecular dynamics framework to account for the quantum-statistical effects of high-frequency vibrations. Excellent agreement with experiments under both isobaric and isochoric conditions within a wide range of temperatures is achieved using our approach. 
\end{abstract}

\maketitle

\section{Introduction\label{intro}}

Heat transport in fluids involves both interatomic interactions and diffusion, and classical \gls{md} simulation is a viable method for computing heat conductivity by including both the interaction and diffusion contributions. Extensive \gls{md} simulations \cite{Bresme2001jcp,zhang2005jpcb,Muscatello2011jcp,Frank2012JCP,Song2014TF,Song2019TF,Gittus2021JCP} have been performed to calculate the heat conductivity of water using various empirical force fields, such as SPC/E \cite{WaterModelSPCE}, TIP4P \cite{Jorgensen1983JCP}, TIP4P/2005 \cite{WaterModelT4P2005}, and ReaxFF \cite{zhang2017jpcb}. However, The force fields were found to have a major influence on the calculated thermal conductivity and quantitative agreement between simulations and experimental measurements in a wide range of temperatures has not been achieved for any force field so far. 

One of the reasons for the disagreement between computations and measurements is the inaccuracy of the empirical force fields. Although classical \gls{md} simulations of heat transport can also be driven by interactions computed by quantum-mechanical \gls{dft} \cite{Marcolongo2016np,kang2017prb}, this approach is currently not efficient enough and has not been extensively applied to heat transport in water. Recently, \glspl{mlp} \cite{behler2007prl} have emerged as an alternative that can achieve the accuracy of \gls{dft} with a small fraction of the cost. Recent studies \cite{morawietz2016pnas,Cheng2019PNAS,monserrat2020nc,Wohlfahrt2020jcp,Zhang2021PRL} have demonstrated the high accuracy of \glspl{mlp} in modeling the thermodynamics of water in various phases. The linear-scaling computational cost with respect to the number of atoms enabled efficient \gls{md} simulation of heat transport in complex systems  that is beyond the reach of perturbative methods \cite{Sosso2012prb,gu2019cms,Mangold2020jap,ying2023ijhmt,dong2023ijhmt,wang2023prb}. A \gls{dp} model \cite{Tisi2021PRB} has been developed to calculate the heat conductivity of water in a range of temperatures. However, only a qualitative agreement with experiments has been achieved. It has been not clear if \glspl{mlp} can reliably predict the heat conductivity of liquid water at a wide range of thermodynamic conditions.

In this work, we developed a \gls{mlp} for water within the \gls{nep} framework \cite{fan2021neuroevolution,fan2022jpcm,fan2022jcp}, which is an efficient \gls{mlp} framework that has been developed with a particular emphasis on heat transport applications. The accuracy of the developed \gls{nep} model is demonstrated by the radial and angular distribution functions as compared to \gls{dft} results. We performed \gls{emd} simulations to calculate the heat conductivity using the Green-Kubo relation \cite{Green1954,Kubo1957}. The results from classical \gls{md} simulations driven by \gls{nep} do not match experiments quantitatively. However, by applying a quantum-statistical correction based on the spectral heat conductivity computed within the \gls{hnemd} approach \cite{fan2018hnemd}, a quantitative agreement with experiments can be achieved for a wide range of temperatures at both isobaric and isochoric conditions. 

\section{\label{met} A NEP model for liquid water  }

The \gls{nep} approach as implemented in the \textsc{gpumd} package \cite{fan2017cpc} has been introduced in Ref. \onlinecite{fan2021neuroevolution} and improved later \cite{fan2022jpcm,fan2022jcp}. This approach follows the work of Behler and Parrinello \cite{behler2007prl} to model the site energy of an atom as an \gls{ann}, where the input layer consists of a descriptor vector of high dimensions. The descriptor components are invariant with respect to the translation, rotation, and permutation of atoms of the same kind. For explicit expressions of the descriptor components, we refer to Ref. \onlinecite{fan2022jcp}. The name \gls{nep} comes from the algorithm for training the \gls{ann}, which is a \gls{snes} \cite{Schaul2011}. 

To train a \gls{nep} model for liquid water, we used the ``refinement'' data set for liquid water taken from Ref.~\onlinecite{Zhang2021PRL}, computed at the quantum-mechanical \gls{dft} level with the \gls{scan} functional \cite{sun2015prl}. There are 1888 structures (each with 128 H$_2$O molecules) in total, and we randomly selected 1388 for training and 500 for testing. For more details on the generation of the reference data, we refer to Ref.~\onlinecite{Zhang2021PRL}.

The hyperparameters we used in the \gls{nep} model are as follows. The \gls{nep} descriptor consists of a number of radial and angular components \cite{fan2021neuroevolution,fan2022jpcm,fan2022jcp}. For the radial components, we used a cutoff radius of 6 \AA{} and ten radial functions (each being a linear combination of 10 basis functions). For the angular components, we used a cutoff radius of 4 \AA{}, eight radial functions (each being a linear combination of 8 basis functions), three-body correlations up to $l=4$ in the spherical harmonics, and four-body correlations up to $l=2$. The \gls{ann} in the \gls{nep} model has a single hidden layer and we used 100 neurons for this layer. 

\begin{figure}[htb]
\begin{center}
\includegraphics[width=1\columnwidth]{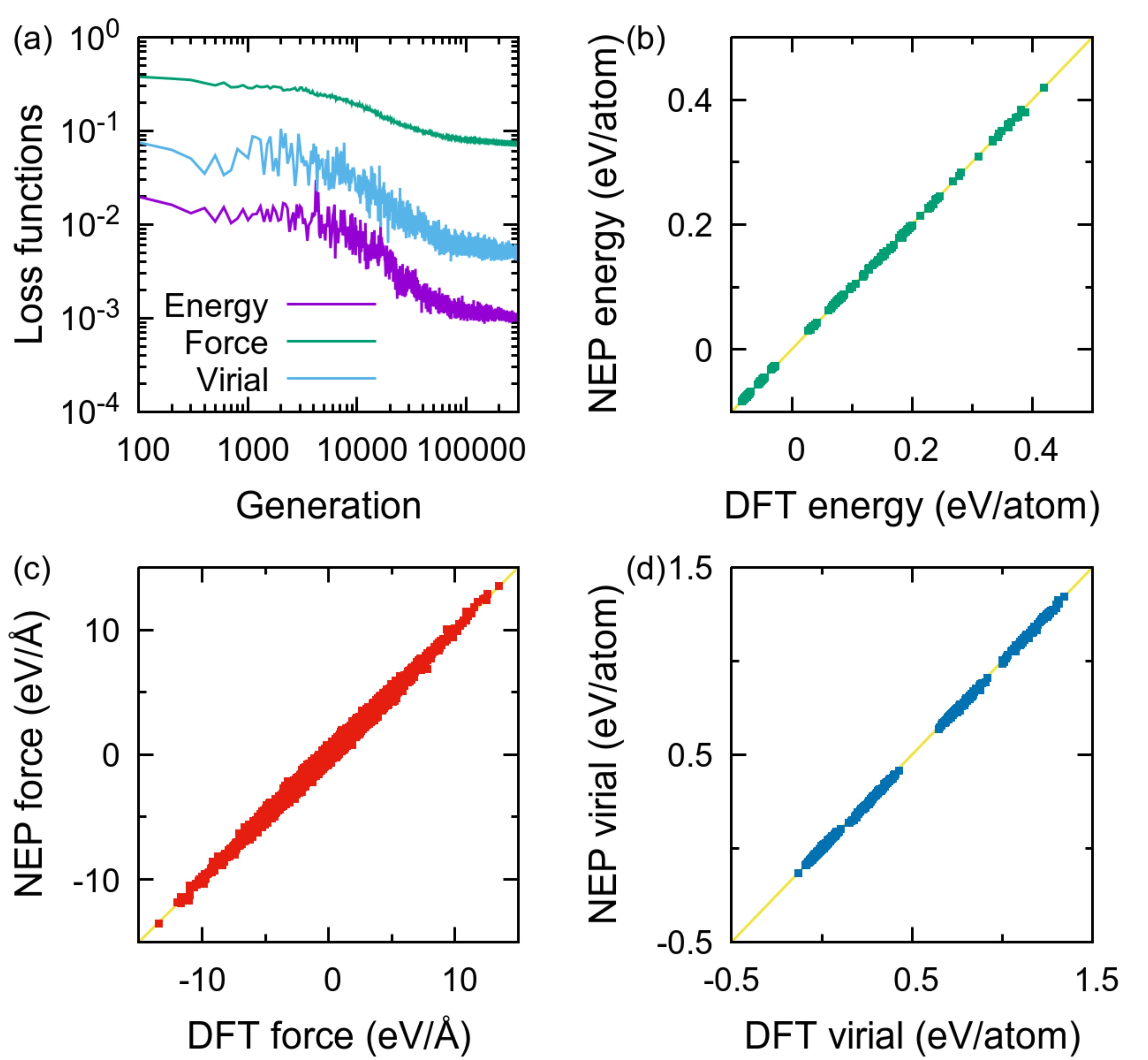}
\caption{\label{fig:neptrain} (a) \glspl{rmse} of energy, force, and virial for the test set as a function of the number of generations. (b)-(d) The comparison between the \gls{nep} predictions and \gls{dft} reference values of energy, force, and virial for the test set. }
\end{center}
\end{figure}

We trained the \gls{nep} model for 300,000 generations using the \gls{snes} algorithm and the loss terms for energy, force, and virial in the test set are largely converged [Fig. \ref{fig:neptrain}(a)]. The predicted energy, force, and virial for the test set are compared to the \gls{dft} reference data in Figs.~\ref{fig:neptrain}(b) to \ref{fig:neptrain}(d), showing good correlations. Quantitatively, the \glspl{rmse} for energy, force, and virial are 0.89 meV/atom, 76 meV/\AA{}, and 5.2 meV/atom in the training data set, and are 1.0 meV/atom, 73 meV/\AA{}, and 5.0 meV/atom in the test data set. The level of accuracy is comparable to those reported in previous works on identical or similar data sets \cite{Zhang2021PRL,Tisi2021PRB}.

\begin{figure}[htb]
\begin{center}
\includegraphics[width=0.9\columnwidth]{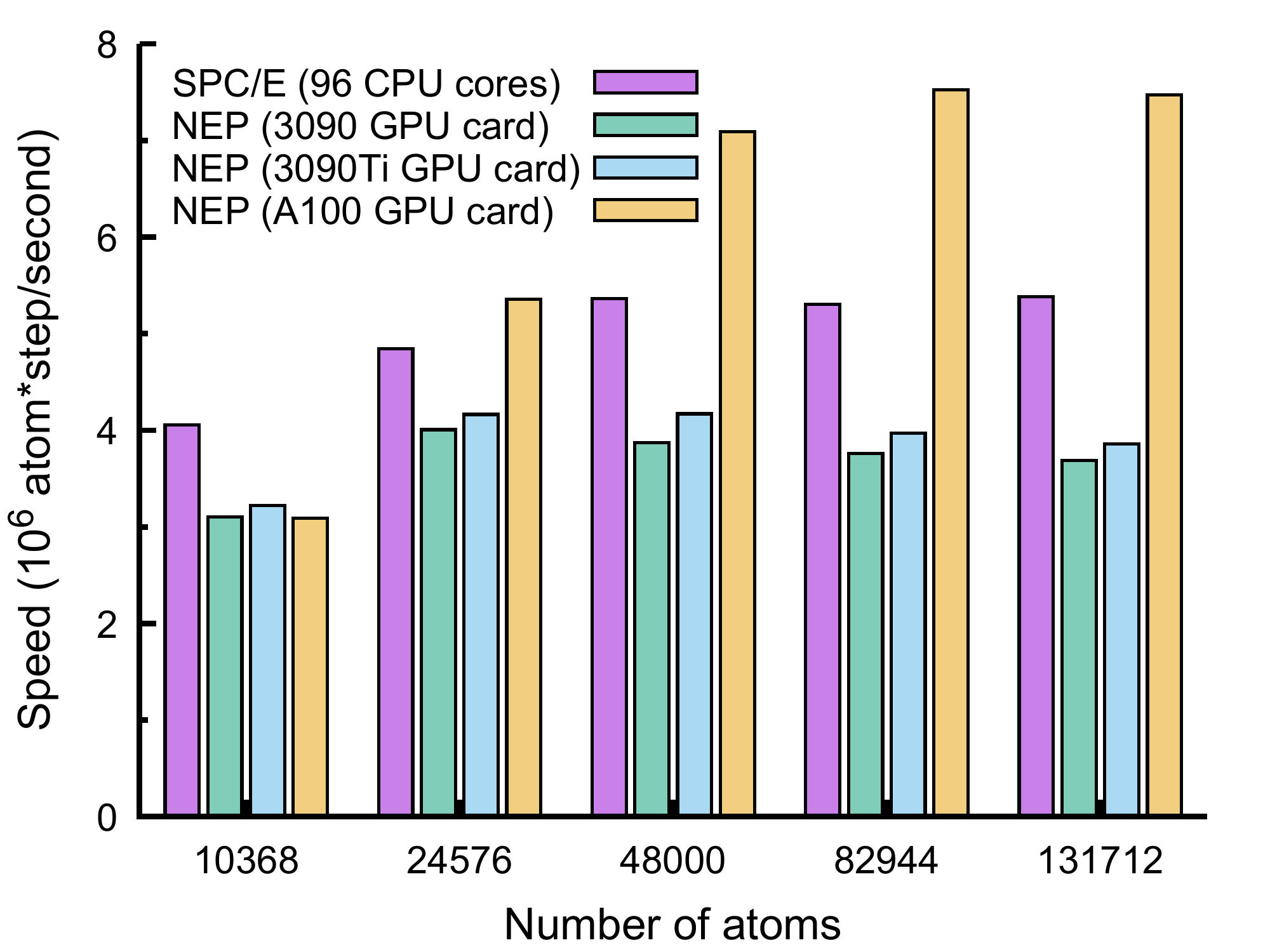}
\caption{\label{fig:speed} Computational speed of our \gls{nep} model as implemented in \textsc{gpumd} \cite{fan2017cpc} in \gls{md} simulations as compared to that of the SPC/E force field as implemented in \textsc{lammps} \cite{thompson2022lammps}. For \gls{nep}, a single GPU (Nvidia 3090, 3090Ti, or A100) was used; for SPC/E, 96 CPU cores (two nodes, each with 48 Intel Xeon Platinum 9242 CPU cores) were used.}
\end{center}
\end{figure}

Our \gls{nep} model can achieve not only a high accuracy but also a high computational speed. To show this, we compare the computational speeds of \gls{nep} as implemented in the \textsc{gpumd} package (version 3.6) \cite{fan2017cpc} and the SPC/E force field as implemented in the \textsc{lammps} package (the 23 Jun 2022 version) \cite{thompson2022lammps}. For the SPC/E force field, the Coulomb interactions were evaluated using the particle-particle particle-mesh (PPPM) method with a real-space cutoff distance of 12 \AA{} and a relative accuracy of $10^{-6}$ in force calculations. Figure~\ref{fig:speed} shows that our \gls{nep} model is literally as fast as the SPC/E force field for comparable amounts of computational resources.      

\begin{figure}[htb]
\begin{center}
\includegraphics[width=1\columnwidth]{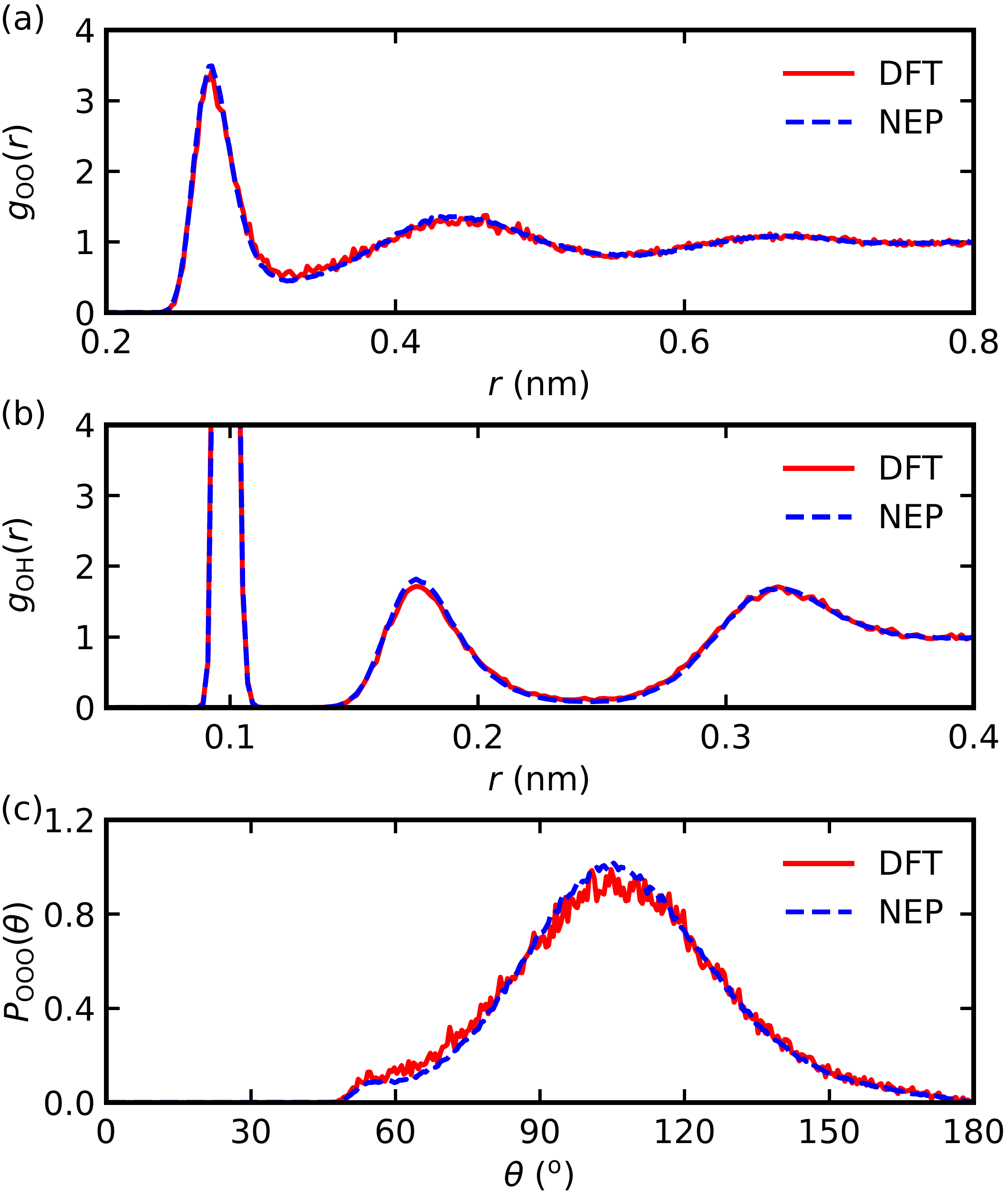}
\caption{\label{fig:rdf} \glspl{rdf} for (a) O-O and (b) O-H pairs, and (c) \gls{adf} for O-O-O triplets, calculated using classical \gls{md} simulations at 300 K and 1 bar driven by \gls{dft} (solid lines) and \gls{nep} (dashed lines).}
\end{center}
\end{figure}

To validate the accuracy of the trained \gls{nep} model in \gls{md} simulations, we compare the \gls{rdf} and \gls{adf} obtained by classical \gls{md} simulations driven by the \gls{nep} model and \gls{dft} calculations, both at 300 K and 1 bar. As shown in Fig. \ref{fig:rdf}, good agreement is achieved for the \glspl{rdf} for O-O pairs, $g_{\rm {OO}}(r)$, and O-H pairs, $g_{\rm {OH}}(r)$, and the \gls{adf} for O-O-O triplets $g_{\rm{OOO}}(\theta)$. The \gls{dft} results were obtained using a small cell with 384 atoms, while the \gls{nep} results were obtained using a much larger cell with 3000 atoms, which explains the much smoother distribution functions from the \gls{nep} model. The good agreement here indicates that our \gls{nep} model can accurately reproduce the dynamics of liquid water, which is a prerequisite for the reliable study of heat transport. 

\section{Heat conductivity of liquid water from MD simulations \label{section:method}}

\subsection{Classical heat conductivity of liquid water}

We calculated the heat conductivity of liquid water using the well established Green-Kubo method \cite{Green1954,Kubo1957}, in which the running heat conductivity $\kappa^{\rm total}(t)$ (the meaining of the superscript ``total'' will become clear below) is calculated as a time integral of the \gls{hcacf} $\langle \bm{J}(t) \cdot \bm{J}(0)\rangle$:
\begin{equation}
\label{equation:gk}
\kappa^{\rm total}(t)=\frac{1}{3k_{\rm B}VT^2} \int_0^t dt' \langle \bm{J}(t') \cdot \bm{J}(0)\rangle,
\end{equation}
where $k_{\rm B}$ is Boltzmann's constant, $T$ and $V$ are the temperature and volume of the system, respectively. The heat current $\bm{J}(t)$ is sampled at an equilibrium state. For liquid molecules, the heat current has two contributions: 
\begin{equation}
\label{equation:JpJk}
\bm{J}=\bm{J}^{\rm k}+\bm{J}^{\rm p}.
\end{equation}
The kinetic term (also called convective term) is 
\begin{equation}
\label{equation:Jk}
\bm{J}^{\rm k} =\sum_{i}{\bm{v}_i E_i},
\end{equation}
and the potential term for many-body potentials such as our \gls{nep} model is \cite{fan2015prb}
\begin{equation}
\label{equation:Jp}
\bm{J}^{\rm p} = \sum_{i} \sum_{j} \bm{r}_{ij} \frac{\partial U_j}{\partial \bm{r}_{ji}} \cdot \bm{v}_i.
\end{equation}
Here, $E_i=\frac{1}{2}m_i\bm{v}_i^2+U_i$ is the total energy of atom $i$, where $m_i$, $\bm{v}_i$, and $U_i$ are respectively the mass, velocity, and potential energy of atom $i$.
According to the decomposition of the heat current, the heat conductivity can be decomposed into three terms: 
\begin{equation}
\kappa^{\rm total}(t) = \kappa^{\rm pp}(t) + \kappa^{\rm kk}(t) + \kappa^{\rm pk}(t),
\end{equation}
where the potential-potential term $\kappa^{\rm pp}$, the kinetic-kinetic term $\kappa^{\rm kk}$, and the cross term $\kappa^{\rm pk}$ correspond to the following \glspl{hcacf}: $\langle \bm{J}^{\rm p}(t) \cdot \bm{J}^{\rm p}(0)\rangle$, $\langle \bm{J}^{\rm k}(t) \cdot \bm{J}^{\rm k}(0)\rangle$, and $\langle \bm{J}^{\rm p}(t) \cdot \bm{J}^{\rm k}(0)\rangle+\langle \bm{J}^{\rm k}(t) \cdot \bm{J}^{\rm p}(0)\rangle$.

The Green-Kubo method is based on \gls{emd}, where the system is first equilibrated in the $NVT$ (constant number of atoms $N$, constant volume $V$, and constant target temperature $T$) or $NpT$ (constant target pressure $p$) ensemble to reach an equilibrium state and the heat currents are then sampled in the $NVE$ (constant energy $E$) ensemble. In all the \gls{md} simulations in this work, we used a time step of 0.1 fs, which has been tested to be small enough. In the \gls{emd} simulations, we used an equilibration time of 50 ps and a production time of 10 ps. For each thermodynamic state with a given temperature and pressure (or density), we performed about 50 independent runs and calculated the statistical error as the standard error between the independent runs. We have tested the effects of finite simulation cells and found that the heat conductivity is essentially unchanged when the linear size of a cubic cell increases from 3 to 11 nm. We chose to use a cell with a linear size of about 6 nm containing 24576 atoms (8192 water molecules) for all the subsequent calculations. 

\begin{figure}[htb]
\begin{center}
\includegraphics[width=1.0\columnwidth]{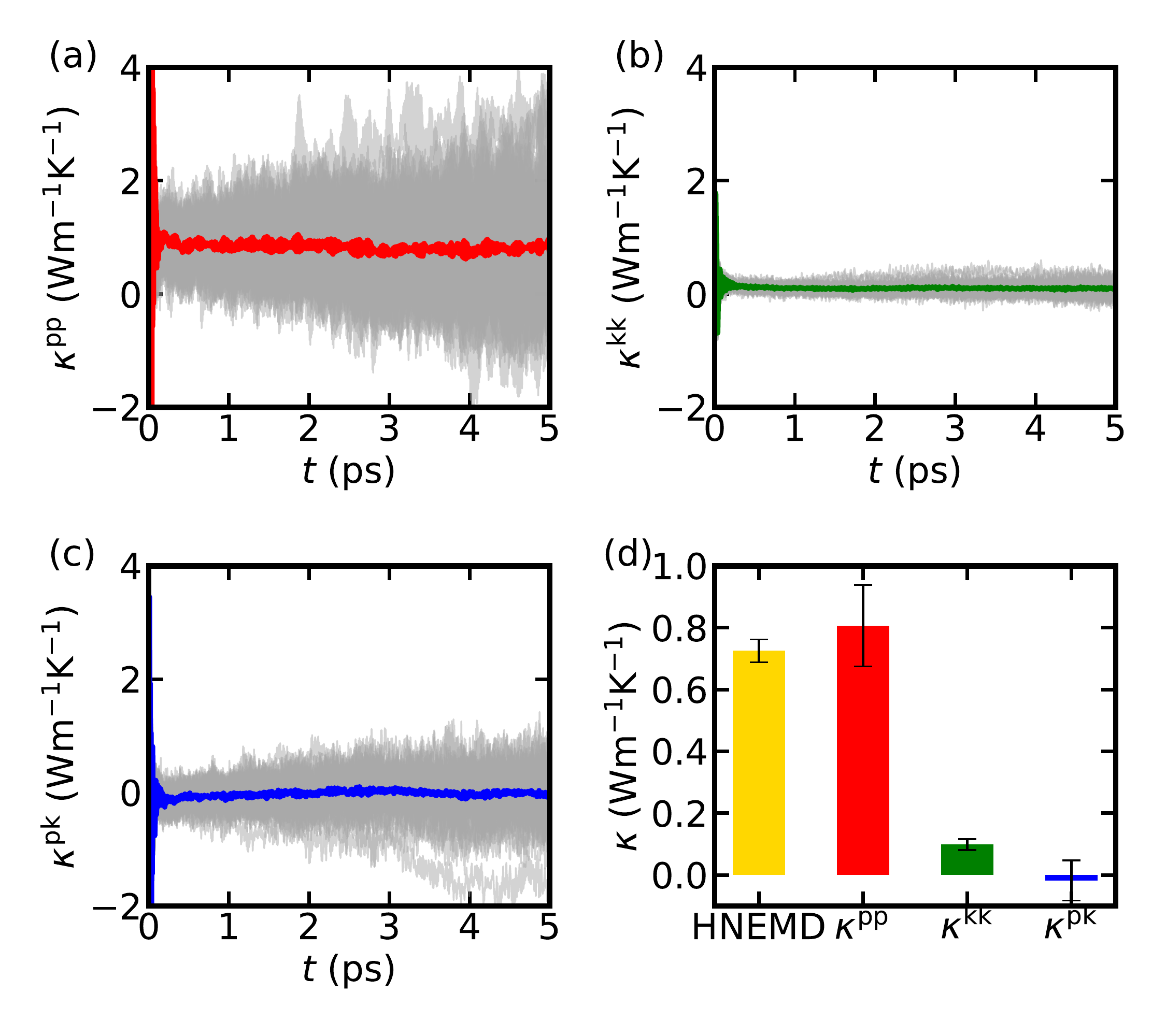}
\caption{\label{fig:EMD} (a)-(c) Running heat conductivity of liquid water at 300 K and 30 bar as a function of correlation time $t$ for the potential-potential term ($\kappa^{\rm pp}$), the kinetic-kinetic term ($\kappa^{\rm kk}$), and the cross term ($\kappa^{\rm pk}$) respectively. (d) Time-converged values for the three components from \gls{emd} simulations and the potential-potential term from \gls{hnemd} simulations. The thin gray lines in (a)-(c) represent results from independent runs, while the thick lines are the averages. }
\end{center}
\end{figure}

Figures~\ref{fig:EMD}(a)-\ref{fig:EMD}(c) show the running heat conductivity components $\kappa^{\rm pp}(t)$, $\kappa^{\rm kk}(t)$, and $\kappa^{\rm pk}(t)$, respectively. In the interval from $t=3$ to 5 ps, all the components show stable oscillations only, without an overall increasing or decreasing trend. We therefore average the running heat conductivity over this time interval for each independent run. With about 50 independent runs, we thus obtained a mean value of each heat conductivity component and a statistical error estimate. These are shown in Fig.~\ref{fig:EMD}(d). Among the three components, $\kappa^{\rm kk}$ is about one order of magnitude smaller than $\kappa^{\rm pp}$ and $\kappa^{\rm pk}$ is essentially zero.

\begin{figure}[htb]
\begin{center}
\includegraphics[width=1\columnwidth]{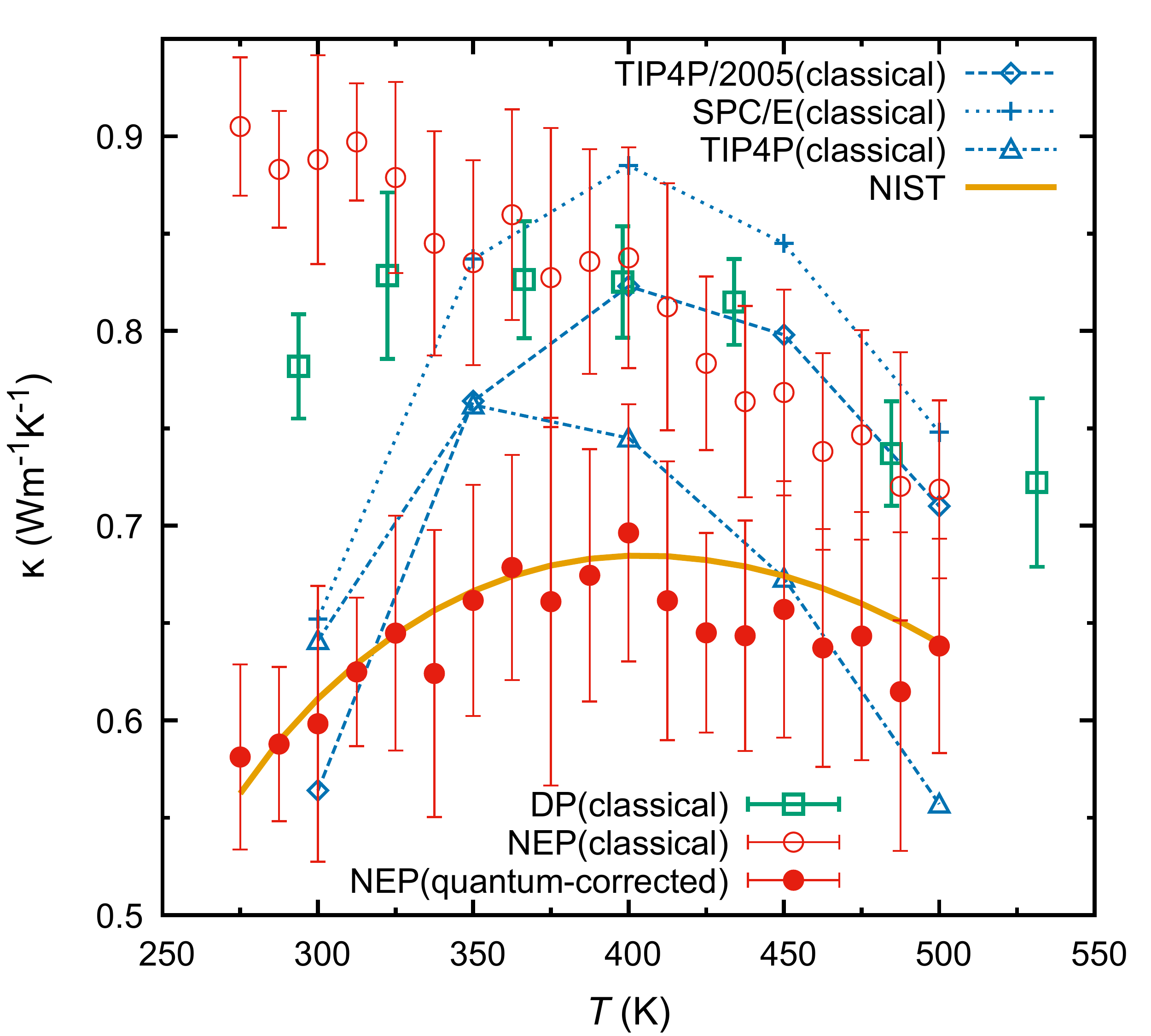}
\caption{Heat conductivity $\kappa$ as a function of temperature $T$ for liquid water from our calculations and previous predictions using the \gls{dp} model \cite{Tisi2021PRB} and three empirical force fields (SPC/E, TIP4P, and TIP4P/2005) \cite{Song2019TF}, and the experimental data from \gls{nist} \cite{Nist_chemistry_webbook,Huber2012JPCRD}. }
\label{figure:kappa_isobaric}
\end{center}
\end{figure}

\begin{figure}[htb]
\begin{center}
\includegraphics[width=\columnwidth]{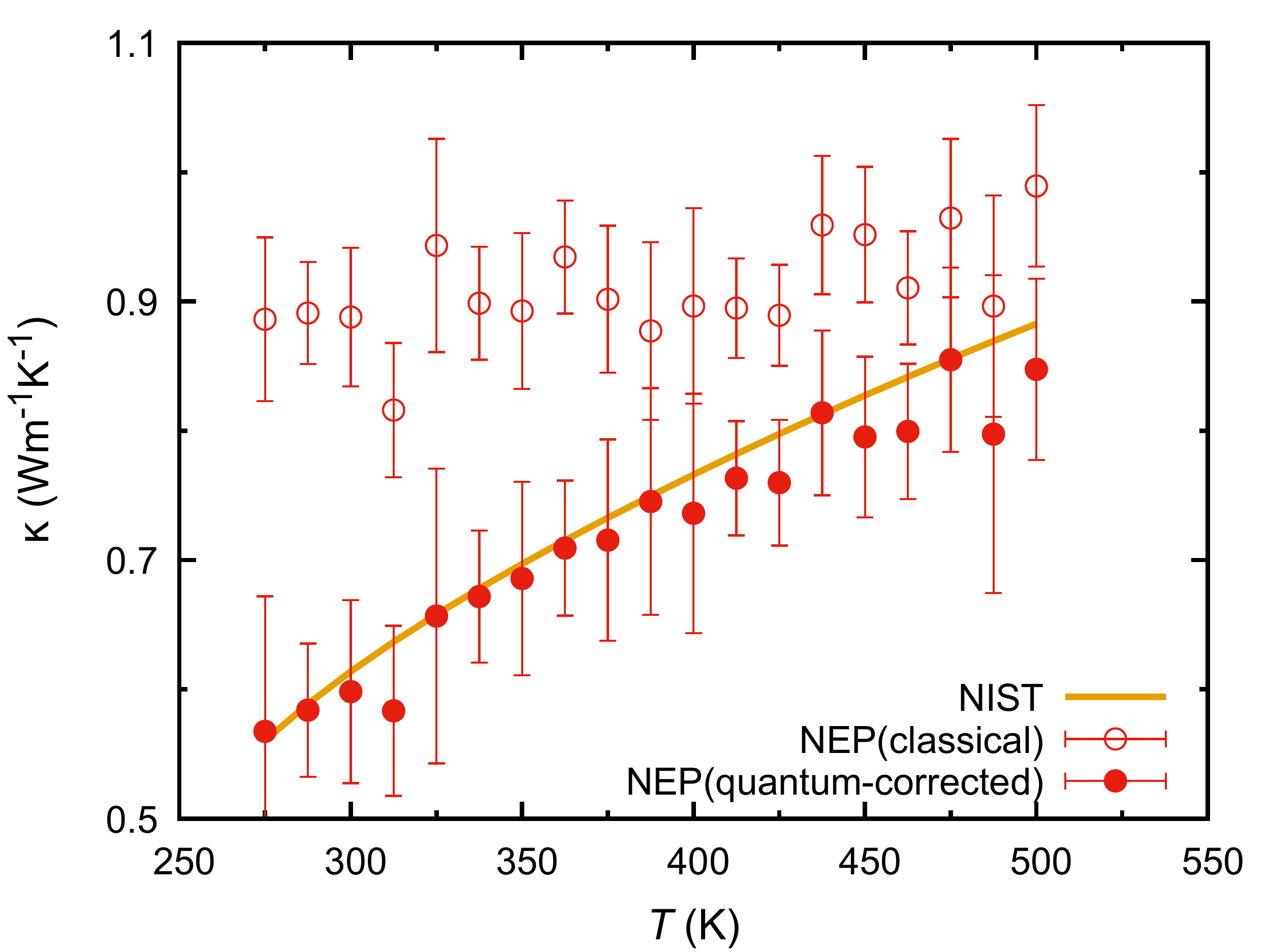}
\caption{Classical and quantum corrected heat conductivity $\kappa$ of liquid water as a function of temperature in the isochoric condition as compared to the experimental values from \gls{nist} \cite{Nist_chemistry_webbook,Huber2012JPCRD}. }
\label{figure:kappa_isochoric}
\end{center}
\end{figure}

Using the Green-Kubo method, we computed the total heat conductivity $\kappa^{\rm total}$ of liquid water from 275 to 500 K (with a step of 12.5 K) under both isobaric and  isochoric conditions. Isobaric conditions were achieved by using the $NpT$ ensemble \cite{Bernetti2020jcp} with a target pressure of 30 bar. Isochoric conditions were achieved by using the $NVT$ ensemble \cite{Bussi2007jcp} with a fixed density of 1 g/cm$^3$. Our results for isobaric and isochoric conditions are presented in Figs.~\ref{figure:kappa_isobaric} and ~\ref{figure:kappa_isochoric}, respectively, along with the experimental data from \gls{nist} \cite{Nist_chemistry_webbook,Huber2012JPCRD} and previous theoretical ones. In the isobaric case, our heat conductivity values (the classical ones) are very close to the previous ones obtained by using the \gls{dp} approach \cite{Tisi2021PRB}, except for a noticeable difference around 300 K. However, the heat conductivity values from both \gls{dp} and our \gls{nep} are significantly higher than the experimental ones, particularly at the lower temperatures. The predicted results from the empirical force fields (SPC/E, TIP4P, and TIP4P/2005) \cite{Song2019TF} show a more complex pattern: they are relatively high at about 400 K but can be close to or lower than experimental values at both the low- and high-temperature limits. Because the experimental data from \gls{nist}  \cite{Nist_chemistry_webbook,Huber2012JPCRD} have small uncertainties (a few percent at most), the results in Fig.~\ref{figure:kappa_isobaric} show that all the theoretical predictions do not quantitatively agree with the experiments. As we will argue below, \glspl{nqe} play an important role here.

\subsection{Quantum-corrected heat conductivity of liquid water}

In classical \gls{md} simulations, the vibrations in the system follow the classical statistics, with all degrees of freedom being fully activated regardless of the temperature and frequency. However, according to quantum statistics, high-frequency degrees of freedom are frozen at low temperatures. Quantitatively, a degree of freedom with frequency $\omega$ at temperature $T$ is only activated with the following probability:
\begin{equation}
\label{equation:p}
    p(x) = \frac{x^2e^x}{(e^x-1)^2},
\end{equation}
where $x=\hbar \omega / k_{\rm{B}} T $, $\hbar$ being the reduced Planck constant. Frequency domain quantum correction \cite{Berens1983JCP} based on the \gls{vdos} has been successfully applied to correct thermodynamic quantities (such as heat capacity) calculated using classical \gls{md}. To our best knowledge, this type of quantum correction has not been applied to heat transport in liquid water. Similar to the quantum correction of heat capacity in water \cite{Berens1983JCP} based on spectral analysis, heat conductivity can be quantum-corrected based on a spectral heat conductivity, as has been recently demonstrated for amorphous silicon in the context of \gls{mlp} \cite{wang2023prb}. Such spectral heat conductivity can be conveniently obtained in the framework of the \gls{hnemd} method as developed in Ref.~\onlinecite{fan2018hnemd}. 

In the \gls{hnemd} simulations, a driving force 
\begin{equation}
\bm{F}_{i} = \bm{F}_{\rm e} \cdot \sum_{j\neq i} \bm{r}_{ij} \otimes \frac{\partial U_j}{\partial \bm{r}_{ji}}
\end{equation}
was applied to each atom $i$ of the system to drive the system into a nonequilibrium steady state, in which the classical spectral heat conductivity is calculated. The vector $\bm{F}_{\rm e}$ represents the driving force parameter that is of the dimension of inverse length. For more details on the \gls{hnemd} method for many-body potentials, we refer to Ref.~\onlinecite{fan2018hnemd}. Five independent runs, each with a production time of 100 ps, were performed to calculate the mean value and statistical error of the heat conductivity. As only the potential-potential part of the \gls{hcacf} involves high frequencies that require a quantum correction, here we only apply the \gls{hnemd} method to calculate the spectral heat conductivity  $\kappa^{\rm pp}(\omega)$, which can be expressed as \cite{fan2018hnemd}:
\begin{equation}
\label{equation:kw}
    \kappa^{\rm pp}(\omega) = \frac{2}{VTF_{\rm e}}\int_{-\infty}^{\infty} \text{d}t e^{\text{i}\omega t} \sum_i\sum_{j\neq i} \left\langle x_{ij} \frac{\partial U_j}{\partial \bm{r}_{ji}}(0) \cdot \bm{v}_i(t) \right\rangle.
\end{equation}
Here we have assumed that heat transport is along the $x$ direction. The magnitude $F_{\rm e}$ of the driving force parameter is set to $0.001$ \AA{}$^{-1}$, which is sufficiently small to keep the system within the linear-response regime.

Eq.~(\ref{equation:kw}) represents a Fourier transform in which the integral is formally from $-\infty$ to $\infty$. In numerical calculations, it is evaluated based on discrete Fourier transform (or more exactly, discrete cosine transform). In \gls{md} simulation, the virial-velocity time correlation function $\langle \cdots \rangle$ in Eq.~(\ref{equation:kw}) is evaluated at discrete times and only up to a finite upper limit $t_{\rm max}$. A Hann window function is applied before performing the discrete cosine transform. According to Nyquist sampling theorem, $t_{\rm max}$ determines the frequency resolution that can be achieved: a larger value of $t_{\rm max}$ results in a finer frequency resolution. In our calculations, we used $t_{\rm max}=250$ fs, which gives a frequency resolution of $1/2t_{\rm max}$ = 2 THz. This is sufficient for our purpose. Using larger $t_{\rm max}$ does not affect any of our results significantly.

\begin{figure}[htb]
\begin{center}
\includegraphics[width=1.0\columnwidth]{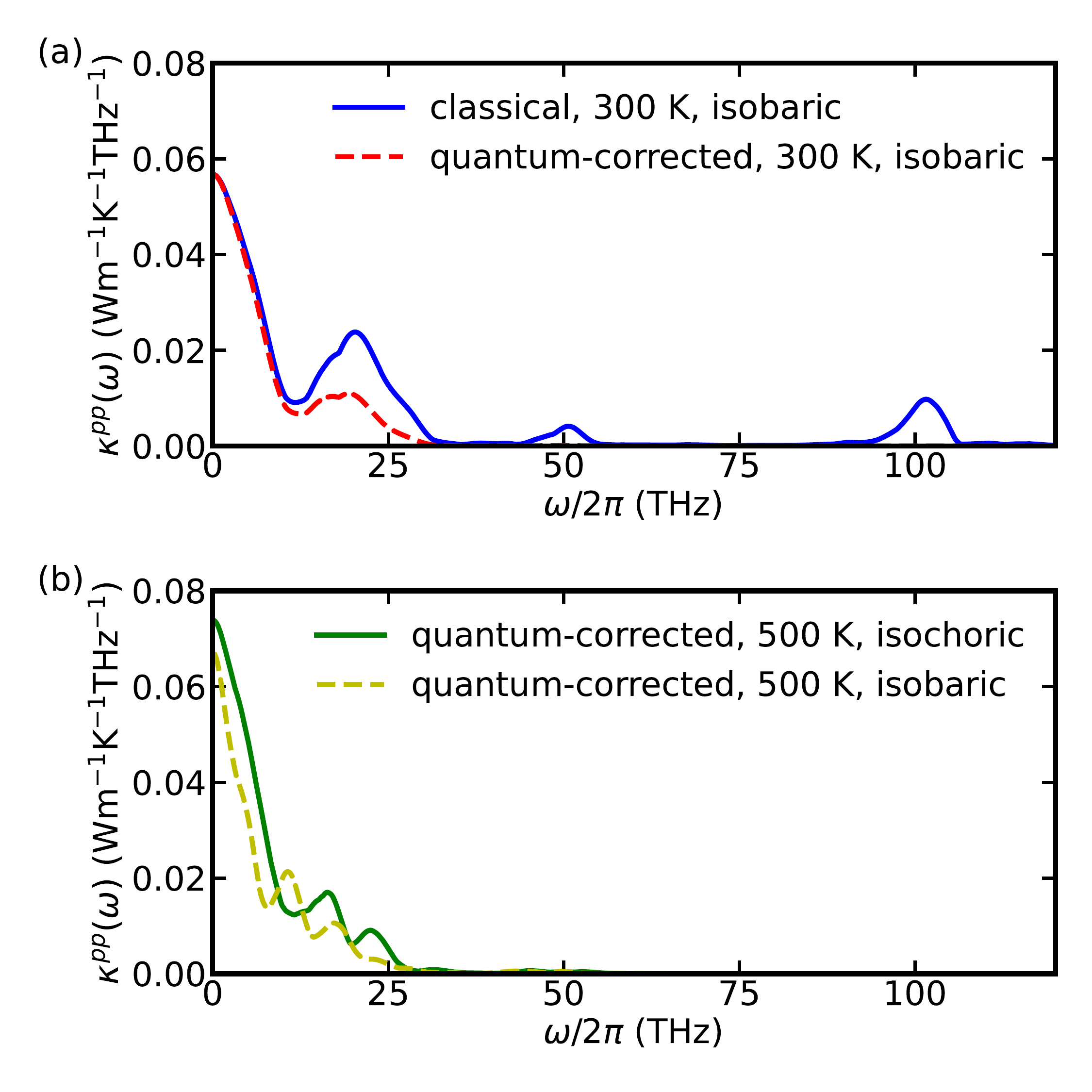}
\caption{(a) Classical and quantum-corrected spectral heat conductivity for liquid water at 300 K and 30 bar. (b) Quantum-corrected spectral heat conductivity for liquid water at 500 K and two different conditions: the isobaric condition with a pressure of 30 bar and the isochoric condition with a density of 1 g/cm$^3$.}
\label{fig:HNEMD}
\end{center}
\end{figure}

The integration of the spectral heat conductivity over the frequency $\omega$ is $\kappa^{\rm pp}$:
\begin{equation}
\kappa^{\rm pp} = \int_0^{\infty} \frac{\text{d} \omega}{2\pi} \kappa^{\rm pp} (\omega).
\end{equation}
For the potential-potential part, \gls{hnemd} and \gls{emd} give consistent heat conductivity, as shown in Fig.~\ref{fig:EMD}(d).
With the classical spectral heat conductivity available, we can then obtain a quantum-corrected spectral heat conductivity $\kappa_{\rm q}^{\rm pp} (\omega)$ by multiplying $\kappa^{\rm pp} (\omega)$ with the probability $p(x)$: 
\begin{equation}
\label{equation:kq}
\kappa^{\rm pp}_{\rm q}(\omega) = \kappa^{\rm pp}(\omega) p(x).
\end{equation}
Figure \ref{fig:HNEMD}(a) shows that the quantum correction is significant at 300 K.

\begin{table}[thb]
\setlength{\tabcolsep}{2Mm}
\caption{Calculated classical and quantum-corrected heat conductivity values (in units of Wm$^{-1}$K$^{-1}$) at different temperatures $T$ (in units of K) under both isobaric and isochoric conditions. The numbers within the parentheses are statistical uncertainties for the last significant number(s).}
\label{table:data}
\begin{tabular}{lllll}
\hline
\hline
& \multicolumn{2}{c}{isobaric} & \multicolumn{2}{c}{isochoric}\\
$T$ & classical & quantum & classical & quantum \\
\hline
275   & 0.91(4) & 0.58(5) & 0.89(6) & 0.57(10) \\
287.5 & 0.88(3) & 0.59(4) & 0.89(4) & 0.58(5) \\
300   & 0.89(5) & 0.60(7) & 0.89(5) & 0.60(7) \\
312.5 & 0.90(3) & 0.62(4) & 0.82(5) & 0.58(7) \\
325   & 0.88(5) & 0.64(6) & 0.94(8) & 0.66(11) \\
337.5 & 0.85(6) & 0.62(7) & 0.90(4) & 0.67(5) \\
350   & 0.84(5) & 0.66(6) & 0.89(6) & 0.69(7) \\
362.5 & 0.86(5) & 0.68(6) & 0.93(4) & 0.71(5) \\
375   & 0.83(8) & 0.66(9) & 0.90(6) & 0.72(8) \\
387.5 & 0.84(6) & 0.67(6) & 0.88(7) & 0.75(9) \\
400   & 0.84(6) & 0.70(7) & 0.90(8) & 0.74(9) \\
412.5 & 0.81(6) & 0.66(7) & 0.89(4) & 0.76(4) \\
425   & 0.78(4) & 0.64(5) & 0.89(4) & 0.76(5) \\
437.5 & 0.76(5) & 0.64(6) & 0.96(5) & 0.81(6) \\
450   & 0.77(5) & 0.66(7) & 0.95(5) & 0.80(6) \\
462.5 & 0.74(5) & 0.64(6) & 0.91(4) & 0.80(5) \\
475   & 0.75(5) & 0.64(6) & 0.96(6) & 0.85(7) \\
487.5 & 0.72(7) & 0.61(8) & 0.90(9) & 0.80(12) \\
500   & 0.72(5) & 0.64(6) & 0.99(6) & 0.85(7) \\
\hline
\hline
\end{tabular}
\end{table}

After applying this quantum correction, we can obtain the overall quantum-corrected potential-potential part of the heat conductivity as
\begin{equation}
\kappa_{\rm q}^{\rm pp} = \int_0^{\infty} \frac{\text{d} \omega}{2\pi} \kappa_{\rm q}^{\rm pp} (\omega).
\end{equation}
Because the kinetic-kinetic and potential-kinetic parts do not involve high-frequency vibrations, they do not need to be quantum corrected. Therefore, we can obtain the quantum-corrected total heat conductivity as
\begin{equation}
    \kappa^{\rm total}_{\rm q} = \kappa^{\rm pp}_{\rm q} + \kappa^{\rm kk} + \kappa^{\rm pk}.
\end{equation}
The total heat conductivity values before and after the quantum correction are listed in Table~\ref{table:data} and are also shown in Figs.~\ref{figure:kappa_isobaric} and \ref{figure:kappa_isochoric}.  

The quantum-corrected total heat conductivity values agree excellently with experiments in the whole temperature range for both isobaric (Fig.~\ref{figure:kappa_isobaric}) and isochoric  (Fig.~\ref{figure:kappa_isochoric}) conditions. Quantitative agreement with experiments for the different temperature and pressure (density) conditions cannot be an accident and it strongly suggests the reliability of our \gls{nep} model and the effectiveness of the quantum correction based on the spectral heat conductivity. Particularly, our approach has correctly predicted the much larger heat conductivity under the isochoric condition than that under the isobaric condition at 500 K. This is intuitively understandable as the density for the isochoric condition is higher than that for the isobaric condition, resulting in stronger interatomic interactions that can enhance the potential-potential part of the heat conductivity. Figure~\ref{fig:HNEMD}(b) further shows that this enhancement mainly comes from the vibrations with $\omega/2\pi < 10$ THz. 

We can now better interpret the theoretical predictions from the \gls{dp} \cite{Tisi2021PRB} model and the three empirical force fields \cite{Song2019TF} as shown in Fig.~\ref{figure:kappa_isobaric}. If the quantum correction were also applied to these predictions, we expect that the \gls{dp} results would agree well with experiments as well, except for the temperatures close to 300 K. On the other hand, all the three empirical force fields would significantly underestimate the experimental results at 300 K. 

\section{Discussion}

Good agreement between our theoretical calculations and experiments clearly depends on both the accuracy of the \gls{nep} model and the effectiveness of the \gls{hnemd}-based quantum correction method. Here we discuss the rationales of the \gls{hnemd}-based quantum correction method and related approaches. 

We start our discussion by presenting a general expression of the spectral heat conductivity: 
\begin{equation}
\label{equation:k_general}
\kappa(\omega) =  c(\omega) v^2(\omega) \tau(\omega),
\end{equation}
where $c(\omega)$ is the modal heat capacity, $v(\omega)$ is the modal group velocity, and $\tau(\omega)$ is the relaxation time of the heat carriers, which are not necessarily phonons but are related to the collective vibrations in the system. 

Clearly, one of the \glspl{nqe} is related to the modal heat capacity: classical \gls{md} overestimates the modal heat capacity by exciting any vibrational mode regardless of its frequency and temperature. The quantum correction for this is simple, which is to multiply $c(\omega)$ by $p(x)$ as defined in Eq.~(\ref{equation:p}). 

There are essentially no \glspl{nqe} in the group velocity but there can be complicated \glspl{nqe} in the relaxation time. This is the case for crystals as has been discussed in the context of phonon Boltzmann transport equation \cite{Turney2009prb,he2012pccp,Puligheddu2019prm}. For example, if classical \gls{md} overestimates the population of a given phonon frequency, it leads to overestimated scattering to other phonons. Whether classical \gls{md} leads to overestimated or underestimated heat conductivity thus depends on the competition between the \glspl{nqe} on $c(\omega)$ and $\tau(\omega)$. Even though classical \gls{md} might lead to the correct total heat conductivity, the spectral heat conductivity can significantly deviate from the quantum result. Therefore, there is so far no feasible quantum-correction method for heat conductivity of crystals for which phonon-phonon scattering is the major source of resistivity \cite{gu2021jap}. Particularly, the temperature-rescaling method \cite{wang1990prb,lee1991prb,volz2000prb} based on equating the classical and quantum energies has been shown to be infeasible \cite{Turney2009prb,Puligheddu2019prm}. 

The situation is different for disordered materials, where the elastic scattering for the vibrational modes by disorder dominates and the population of vibrations has negligible effects on the elastic scattering processes. Therefore, the major quantum effects in classical simulation of disordered systems are from the overestimated modal heat capacity. In this case, the spectral heat conductivity can be quantum corrected by multiplying it with $p(x)$, which is consistent with the \gls{hnemd}-based quantum-correction method. The effectiveness of this quantum-correction method has been recently demonstrated for amorphous materials \cite{wang2023prb} and our current work extends its applicability to liquids by complementing it with \gls{emd} simulations for convective heat transport.

While we have only studied liquid water in this work, we believe that our approach is also applicable to other fluids with light elements at relatively low temperatures. However, a more systematic study is needed to evaluate the effectiveness of our approach in other systems. We note that quantum \gls{md} methods such as linearized semiclassical initial value representation, centroid \gls{md}, and ring-polymer \gls{md} have been used to study heat transport of both liquids \cite{Yonetani2003jcp,liu2011jcp,Sutherland2021jcp} and solids \cite{luo2020jcp} to account for the \glspl{nqe}. The relative performance of our approach compared to these quantum \gls{md} methods in predicting heat conductivity remains to be explored. Particularly, our approach does not account for \glspl{nqe} in $\kappa^{\rm kk}$, which is essentially a zero-frequency property similar to the diffusion coefficient. As it has been shown that there are large \glspl{nqe} in the diffusion coefficient of liquid water \cite{Marsalek2017jpcl}, we expect that there are also \glspl{nqe} in $\kappa^{\rm kk}$. However, we note that the classical value of $\kappa^{\rm kk}$ only contributes about 10\% to the total heat conductivity. Therefore, even though we have not quantum-corrected $\kappa^{\rm kk}$, we have only ignored little \glspl{nqe} for the total heat conductivity.

\section{Conclusions}
In summary, we have constructed a \gls{nep} model for liquid water that can accurately reproduce structural properties as determined by quantum-mechanical \gls{dft} calculations. The \gls{nep} model is as efficient as empirical force fields of liquid water in large-scale \gls{md} simulations. Heat conductivity values calculated using the Green-Kubo method within classical \gls{md} simulations were found to be overestimated against experimental results, particularly for relatively low temperatures. This led us to identify the importance of \glspl{nqe} in determining the heat conductivity of water. We then proposed a scheme of quantum correction based on the spectral heat conductivity as calculated within the framework of \gls{hnemd} simulations, which leads to excellent agreement with experiments under both isobaric and isochoric conditions within a large range of temperatures. 

\begin{acknowledgments}
K.X, Y.H, and J.W acknowledge support from the National Natural Science Foundation of China (NSFC) (No. 12172314, 11772278, and 11904300), the Jiangxi Provincial Outstanding Young Talents Program (No. 20192BCBL23029), the Fundamental Research Funds for the Central Universities (Xiamen University: No. 20720210025), and the 111 project (B16029). Z.F. acknowledges support from NSFC (No. 11974059).
\end{acknowledgments}

\vspace{0.5cm} 

\noindent{\textbf{Conflict of Interest}}

The authors have no conflicts to disclose.

\noindent{\textbf{Data availability}}

The source code and documentation for  \textsc{gpumd} are available
at \url{https://github.com/brucefan1983/GPUMD} and \url{https://gpumd.org}, respectively. The training and testing results for the NEP model are freely available at \url{https://gitlab.com/brucefan1983/nep-data}.

\end{document}